# Directional visible light scattering by silicon nanoparticles

Yuan Hsing Fu,[1] Arseniy I. Kuznetsov,[1,*] Andrey E. Miroshnichenko,[2] Ye Feng Yu,[1] and Boris Luk'yanchuk,[1]

1. Data Storage Institute, 5 Engineering Drive 1, 117608, Singapore

2. Nonlinear Physics Centre, Research School of Physics and Engineering, the Australian National University, Canberra, 0200, Australia

*Corresponding author: arseniy_k@dsi.a-star.edu.sg

**Keywords:** dielectric nanoparticles, nanoantenna, magnetic dipole, visible spectrum.

**Abstract:** Directional light scattering by spherical silicon nanoparticles in the visible spectral range is experimentally demonstrated for the first time. These unique scattering properties arise due to simultaneous excitation and mutual interference of magnetic and electric dipole resonances inside a single nanosphere. Directivity of the far-field radiation pattern can be controlled by changing light wavelength and the nanoparticle size. Forward-to-backward scattering ratio above 6 can be experimentally obtained at visible wavelengths. These unique properties of silicon nanoparticles make them promising for design of novel low-loss visible- and telecom-range nanoantenna devices.

Efficient control of the visible light at the nanoscale dimensions is important for future light-on-chip integration. At the sub-wavelength scale conventional optical elements, such as lenses, are not functional, which requires conceptually new types of nanoscale optical devices. Resonant plasmonic structures are one of the most promising solutions to this problem due to their ability to capture and concentrate visible light at sub-wavelength dimensions.[1, 2] Different types of nanoscale optical devices such as nanolenses,[3, 4] nano-waveguides,[5, 6] nanoantennas,[7-14] etc. based on metallic nanoparticles have been demonstrated. Many of these functional devices such as nanoantennas are designed in analogy with microwave optics, where metals are routinely used for manipulation of the electromagnetic radiation.[8, 9, 12] The main drawback of using plasmonic particles in the visible frequency range is their intrinsic losses, which strongly affect their overall performance and limits their scalability to dimensions for practical use. One of the possible ways to avoid such limitation and still have similar resonant properties, is to use high-refractive index dielectric nanoparticles.[15, 16] Recently, visible-range nanoantennas based on spherical silicon nanoparticles have been theoretically analyzed demonstrating higher efficiency compared to their metallic analogues.[17-19] However, experimental demonstrations of such antennas have been limited to millimeter scales and microwave frequencies.[20]

According to theoretical predictions based on Mie theory high-refractive index dielectric nanoparticles of spherical shape can have strong dipole-like resonances in the visible spectral range.[21, 22] In contrast to spherical metallic nanoparticles, for which the resonant scattering is dominated by the electric-type resonances (electric dipole, quadrupole etc.),[23] dielectric nanoparticles can have both electric and magnetic dipole resonances simultaneously excited inside the same particle.[21, 22, 24] The magnetic dipole response of dielectric particles originates from the circular displacement currents excited inside the particle by incident light. Recently, first experimental demonstrations of a very strong magnetic dipole resonance in the visible spectral range using silicon nanoparticles have been published.[25, 26]

On the other hand, as shown in recent theoretical works, interference of magnetic and electric dipole resonances inside high-refractive index dielectric nanoparticles can strongly affect their scattering pattern making it dependent on wavelength.[27, 28] For some wavelengths the particles can act as "Huygens" sources, scattering the whole energy in the forward direction, while for another wavelength range, light is almost completely scattered backward. This behavior is similar to scattering of hypothetic magneto-dielectric particles theoretically predicted by Kerker et al. in 1983.[29] Such behavior is also associated with the directional Fano resonance,[30-32] when the scattering is strongly enhanced due to constructive resonant interference in one direction and suppressed in the opposite one. It can be achieved e.g. in the system with two dipole-like excitations, one of them being at the resonance.[33]

To date, no experimental proof of this concept in the optical spectral range has been demonstrated. Realization of such relatively simple nano-optical systems having both electric and magnetic dipole resonances with shifted phases would allow scaling the optical nanoantenna concept down to a single nanoparticle, which can redirect incoming light in different directions depending on wavelength and size.[17, 18, 34, 35]

In this paper, we present the first experimental study of directional scattering by spherical silicon nanoparticles in the visible spectral range. We demonstrate strong anisotropy of nanoparticle scattering in forward and backward directions. This leads to the appearance of a large spectral range where forward scattering strongly dominates, exceeding its backward counterpart by more than 6 times.

**Results and Discussion**

Fig. 1 shows scattering properties of a spherical silicon nanoparticle with radius of 75 nm in free space calculated by Mie theory.[36] The particle is excited by a plane wave from top and the scattering into the upper (backward) or lower (forward) hemispheres is calculated as shown in the inset to Fig. 1a. Fig. 1a represents forward (green curve) and backward (blue curve) scattering spectra together with the forward-to-backward ratio calculated dividing one spectrum by another (orange curve). The scattering spectra into forward and backward directions are very different. Both spectra have two well defined maxima in the visible spectral range, which can be attributed to electric (around 500 nm) and magnetic (around 600 nm) dipole resonances.[25] However, one can see a noticeable shift of the directional scattering resonance positions. For backward scattering, the electric and magnetic dipole resonances are located closer to each other than in the case of forward scattering. There are three well defined spectral ranges with different scattering properties. In the first spectral range with wavelengths $\lambda > 603$ nm, forward scattering dominates while backward scattering is almost zero, which gives rise to the forward-to-backward ratio reaching a maximum value of ~8 at 660 nm. The calculated angular scattering diagram at this wavelength is plotted in Fig. 1b and demonstrates almost perfect alignment of the scattering into the forward direction. In the second wavelength range of 500 nm $< \lambda <$ 603 nm the situation completely changes. The backward scattering is dominant while forward-to-backward ratio goes below 0.5 at 564 nm. The scattering diagram at this wavelength (also shown in Fig. 1b) demonstrates the scattering alignment into the backward direction with minor residual scattering in the forward direction. In the third spectral range $\lambda <$ 500 nm the forward-to-backward ratio starts to grow again, which is accompanied by a reduction of total scattering making this spectral range less attractive. Two more interesting spectral points can be noted at 603 nm and 500 nm where the forward and backward scattering intensities are equal. At 603 nm the scattering pattern corresponds to a magnetic dipole aligned in the Y direction parallel to the magnetic field of incoming light, while at 500 nm it corresponds to an electric dipole aligned in the X direction parallel to the electric field (Fig. 1b).

Scattering anisotropy described above has recently been discussed theoretically for silicon[27] and germanium[28] nanoparticles. In particular, it was shown that the wavelengths of the maximum forward and backward scattering correspond well to conditions derived by Kerker et al. for hypothetic magneto-dielectric particles[29]. The reason for anisotropic scattering is the presence of two orthogonal dipole resonances (electric and magnetic), whose interference influences the total scattering pattern. The remainder of this paper will concentrate on experimental demonstration of this concept in the visible spectral range.

In this paper, we used femtosecond laser ablation method to produce silicon nanoparticles of various sizes. In contrast to the previous studies,[25] the ablated silicon wafer was covered by an additional glass substrate. The ablated nanoparticles were collected and solidified on this glass substrate forming silicon nanoparticles of almost perfect spherical shape. This method for particle generation is somewhat similar to laser-induced transfer,[37-40] but with no real control over nanoparticle size and position on the substrate. Fig. 2 shows dark-field microscope images of silicon nanoparticles collected on the glass substrate. Both reflected and transmitted dark-field images have been recorded, showing scattering by these nanoparticles in backward and in forward directions (see experimental geometries in Fig. 2 a&b and Supplementary Information for details). In these images, fabricated nanoparticles shine by all the rainbow colors, which correspond to their strong magnetic and electric dipole scattering.[25] However, the colors of the same particles are systematically different in the forward and in the backward directions (see side-to-side comparison in Fig. 2 c&d). For example, particle #2 has a blue color in the backward scattering direction, and is green in the forward scattering direction. The same is for particle #4, which color changes from green (backward) to yellow (forward). Similar color modifications can be observed for each silicon nanoparticle in Fig. 2. Such color changes in forward and backward directions have earlier been observed in gold-silver dimers exhibiting Fano-type resonances.[33] In the case of metallic dimer, two electric dipoles excited in different nanoparticles interfere providing Fano-type resonant anisotropic scattering. In silicon nanoparticles, both electric and magnetic dipoles are excited simultaneously, which gives rise to anisotropic scattering by a single nanoparticle of spherical shape.

To account for this effect we performed single nanoparticle spectroscopy measurements for each of six nanoparticles marked in Fig. 2 with numbers 1-6 in both forward and backward scattering directions (see details in Methods section). Results of these measurements are shown in Fig. 3 together with the nanoparticle dark-field microscope and SEM images. As it can be seen from the SEM analysis each shiny object in Fig. 2 corresponds to a silicon nanoparticle of shape close to spherical and diameter ranging from 100 nm to 200 nm. However, forward (green curves) and backward (blue curves) scattering spectra are very different from each other for each nanoparticle. The resonance maxima positions in the backward scattering are much closer one to each other, which is similar to theoretically predicted behavior, making them almost indistinguishable in the experiment. On the other hand maxima in the forward scattering spectra are well defined. This provides a large spectral range at the long-wavelength side where forward scattering is strongly dominant over backward. Orange curves in Fig. 3 show experimental data for the forward-to-backward ratio, which was obtained for each nanoparticle dividing one scattering spectrum by another. This ratio has well defined maximum at the long wavelength spectral side and reaches value above 6 for some nanoparticle sizes (see the right axes in Fig. 3). In this spectral range silicon nanoparticles act as "Huygens" sources scattering light only in forward direction. On the other hand in the central spectral range between electric and magnetic dipole resonances the backward scattering becomes dominant. Experimentally measured spectral behavior is very similar to the theoretically predicted one and proves that a single silicon nanoparticle can controllably redirect incoming light in forward or backward direction depending on nanoparticle size and irradiation wavelength.

Experimentally observed spectra are quite similar to theoretical calculations shown in Fig. 1 (see particles #4&5 for close resonance positions). However, one can also see some differences. Electric dipole resonances in the experiment are less pronounced and located closer to magnetic dipole resonance peaks than in the theory which makes them almost indistinguishable in the backward scattering measurements. We hypothesize this difference arises from non-ideality of the nanoparticle shape. The SEM analysis (see insets in Fig. 3) shows that almost all nanoparticles are slightly squeezed in vertical direction. To account for this effect we calculated spectral properties of nanoparticles with spheroidal shape with X and Y radii ($R_x$ and $R_y$) fixed at 75 nm and radius in Z direction ($R_z$) varied from 75 nm to 40 nm (see Fig. 4). These calculations were performed using finite-difference time-domain (FDTD) method by commercial FDTD software (Lumerical Solutions, Inc.) in the same geometry as in Fig. 1. When the particle is squeezed in Z direction the scattering electric and magnetic resonances become less pronounced and shift closer to each other. This leads to overlapping of the resonances in the backward scattering at relatively large aspect ratios ($R_z < 60$ nm). This spectral behavior is similar to that experimentally observed for particle #4, which has slightly spheroidal shape (see Fig. 3 (#4) and Fig. 4d for direct comparison). On the other hand particle #5 has a shape much closer to a sphere, which results in better defined electric and magnetic dipole resonances in both forward and backward scattering directions. Another interesting feature arising from these calculations is that the maximum of the forward-to-backward ratio shifts to shorter wavelengths and approaches the maximum of forward scattering when the particle is squeezed in Z direction. This results from spectral overlap of electric and magnetic dipole resonances and leads to an increase of the total scattering in the forward direction. Similar effects were recently theoretically predicted for silver core/silicon shell nanoparticles where overlapping of electric and magnetic resonance was achieved by changing the size of the silver core.[19]

Further deviations between theory and experiment arise due to the glass substrate, which effect shifts electric dipole resonance to the longer wavelengths closer to magnetic dipole resonance due to "dressing effect".[25] This behavior is in good correlation with our experimental results (Fig. 3). We should also note that small difference between experimental nanoparticle sizes obtained from SEM analysis and those predicted theoretically for the same resonance positions can be explained by the presence of a thin natural oxide layer on top of the nanoparticles, which effectively decreases the silicon nanoparticle radius.[26]

In conclusion, it is experimentally demonstrated for the first time that silicon nanoparticles of spherical shape exhibit strongly anisotropic scattering in the visible spectral range. This unique scattering behavior results from interference of electric and magnetic dipoles, excited by external illumination. As a result in a broad spectral range these nanoparticles can scatter the most of energy in the forward direction acting like "Huygens" sources. For a different spectral range the most of the energy is scattered in the backward direction. Control and fine tuning of these spectral regions is possible due to almost linear dependence of the scattering resonances on the nanoparticle size. It is also shown that by squeezing the nanoparticles in the direction of light propagation it is possible to further overlap electric and magnetic dipole resonances. This results in a noticeable increase of their forward scattering.

These unique optical properties and very low losses compared to plasmonic nanoparticles make silicon nanoparticles perfect candidates for design of high-performance nanoantennas, low-loss metamaterials, and other novel nanophotonic devices.

During the preparation of this paper another work has been published,[41] which demonstrates directional scattering of high-refractive index dielectric nanoparticles. These results were obtained in GHz frequency range with millimeter-size particles. Our work presents the first experimental demonstration of Kerker-type scattering at optical frequencies.

**Methods**

Si nanoparticles of various sizes were fabricated by femtosecond laser ablation of a silicon wafer. A glass substrate was placed on top of the wafer. The laser beam passed through the glass substrate and ablated silicon surface. Ablated silicon nanoparticles were collected and solidified on the glass substrate. Then they were studied by single nanoparticle dark-field spectroscopy, dark-field optical microscopy and scanning electron microscopy (DA300, FEI). In the laser experiments, we used a commercial 1 kHz femtosecond laser system (Tsunami+Spitfire, Spectra Physics) delivering 1 mJ, 100 fs laser pulses at a central wavelength of 800 nm. The laser beam with diameter of 4 mm was focused onto the sample surface by a 20x microscope objective (Mitutoyo, MPlan NIR 20). Laser irradiation conditions were chosen to ablate a moderate amount of nanoparticles, high enough to have all the required sizes but small enough to keep individual nanoparticles separated from each other.[25] The laser beam was scanned through the sample surface at average power of 0.25 mW and scanning speed of 1 mm/s. These parameters are slightly above the ablation threshold of silicon at these irradiation conditions. Scanned lines were separated by a pitch of 100 μm to avoid overlapping of ablated particles from different scans. The ablated particles were observed as bright colored spots in a dark-filed microscope (Nikon, Ti-U) at 50x magnification.

Optical scattering properties of single nanoparticles were studied with an optical dark-field microscope (Nikon, Ti-U) equipped with a high-sensitivity spectrometer (Andor SR-303i) and a 400 × 1600 pixel EMCCD (Andor Newton) described elsewhere.[25] Fig. 2a&b show how the forward and backward scattering of single nanoparticles were measured. Si nanoparticles on glass substrate were placed facedown onto the sample holder of the microscope. During forward scattering measurements, the light from the upper dark-field condenser shone onto the sample surface, and the scattering light was collected by the lower dark-field objective lens. During backward scattering measurements, the light shone from the lower dark-field objective lens, and the scattered light was collected by the same lens. The numerical aperture (NA) of the upper dark-field condenser and the lower dark-field objective lens for light incidence was 0.8-0.9, and NA of the objective lens for light collection was 0.55. In these conditions, the most forward scattering and most backward scattering are collected separately (see more detailed scheme in Fig. S1 in Supplementary Information). To normalize forward and backward scattering intensities to the same level, a scattering from a sub-wavelength dust spot was measured in both geometries and used for normalization. A colored camera was also attached to another port of the microscope to collect optical images of the scattering pattern in the dark-field illumination mode. The scattering spectra were measured over the wavelength range of 400-800 nm.

**Acknowledgment**

The authors thank Ms. Janaki DO Shanmugam (DSI) for help with SEM analysis and Reuben Backer (DSI) for editing. This work was supported by the Agency for Science, Technology and Research (A*STAR) of Singapore: SERC Metamaterials Program on Superlens, grant no. 092 154 0099; SERC grant no. TSRP-102 152 0018; the grant no. JCOAG03-FG04-2009 from the Joint Council of A*STAR, and by the Australian Research Council through the Future Fellowship project FT110100037.

**Author contributions**

YHF, AIK, and BL contributed to initial idea generation; YHF performed single nanoparticle spectroscopy and FDTD modeling, contributed to SEM analysis and wrote the first paper draft; AIK fabricated Si nanoparticles, contributed to single nanoparticles spectroscopy and SEM analysis, wrote the final manuscript and coordinated the whole work; AEM performed Mie theory simulations of a Si nanoparticle in free space and theoretical analysis of a Si nanoparticle on a substrate; YFY contributed to discussions and FDTD simulations; BL and AEM contributed to the manuscript preparation. All authors read and corrected the manuscript before the submission.

**Additional information**

Competing financial interests: The authors declare no competing financial interests.

**Figure legends**

**FIGURE 1.** Scattering properties of a silicon (Si) nanoparticle with radius of 75 nm in free space calculated by Mie theory. **(a)** Forward (green curve) and the backward (blue curve) scattering cross-sections, and the forward-to-backward ratio (orange curve) of the Si nanoparticle. The grey dashed line represents the values of forward-to-backward ratio equal to unity. Inset into **(a)** is the calculation scheme: the nanoparticle is excited by a plane wave from top while the scattered light is integrated over the upper and lower hemispheres for backward and forward scattering. **(b)** Far-field scattering radiation patterns in the four spectral points: 660 nm – maximum forward-to-backward ratio; 564 nm – minimum forward-to-backward ratio; 500 nm and 603 nm – equal forward and backward scattering at electric (500 nm) and magnetic (603 nm) dipole resonances.

**FIGURE 2.** Transmission and reflection dark-field microscope images of laser-generated Si nanoparticles. **(a)&(b)** Schemes of forward and backward scattering measurements. **(c)&(d)** True color CCD images of the forward and backward scattering by Si nanoparticles. Selected nanoparticles are marked by corresponding numbers from 1 to 6 in both images. Inset into **(c)&(d)** shows magnified dark-field microscope image of nanoparticle #4 in forward and backward scattering directions.

**FIGURE 3.** Experimentally measured forward (green curves) and backward (blue curves) scattering spectra of silicon nanoparticles selected in Fig. 2. Left axes show forward and backward scattering intensities, and right axes show forward-to-backward ratio (orange curves). Dashed lines represent the forward-to-backward ratio equal to unity. Dots represent experimental data, and solid lines are their computer-generated smoothing. Insets show close view transmitted (F) and reflected (B) dark-field microscope images, and SEM images taken at an angle of 52° for each selected nanoparticle (1–6). The scale bar in the SEM images is 500 nm.

**FIGURE 4** Influence of the shape of Si nanoparticles on their spectral properties. **(a)** Schematic representation of a spheroidal nanoparticle with $R_x = R_y$, and varied $R_z$. $R_x$ and $R_y$ are fixed at 75 nm. **(b)–(f)** Calculated results of forward (green curves) and backward (blue curves) scattering, and forward-to-backward ratio (orange curves) for silicon nanoparticles with $R_z$ = 75, 70, 60, 50, and 40 nm, respectively. Calculations are done in free space using FDTD method.

**Fig. 1**

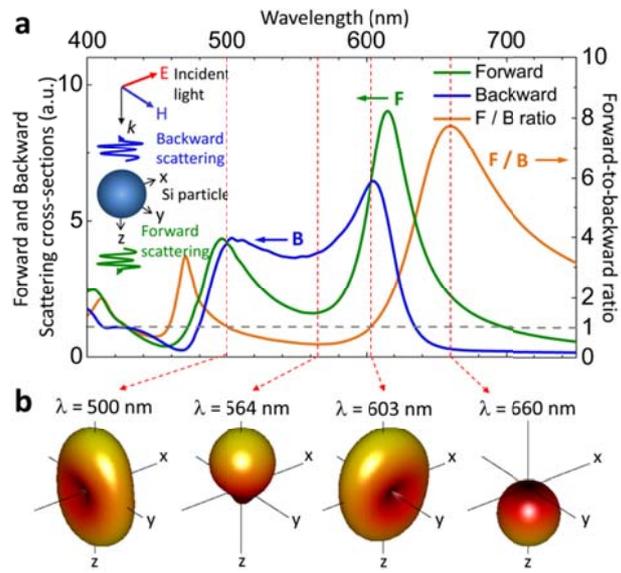

**Fig. 2**

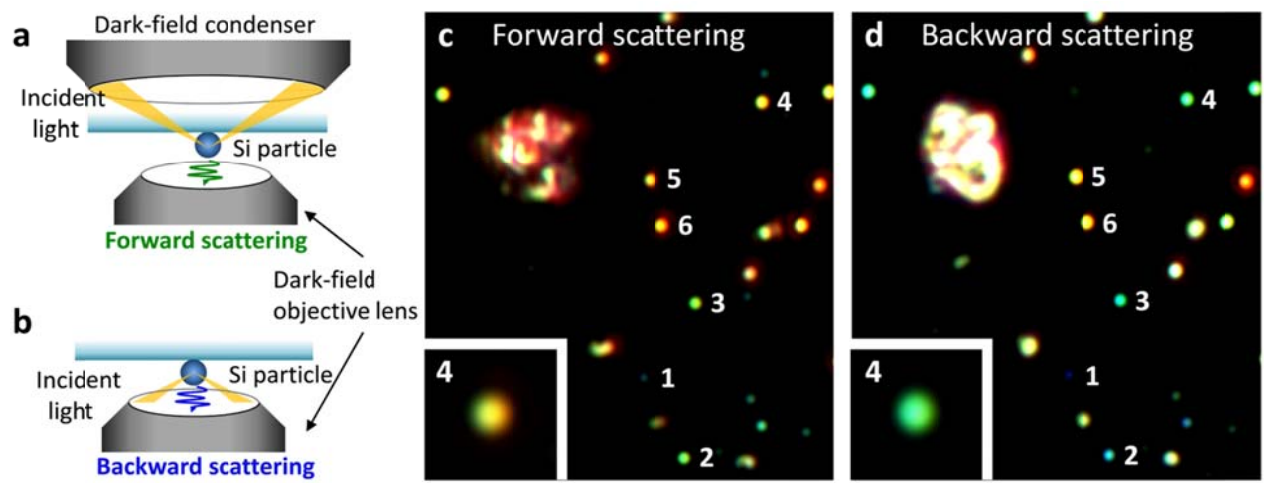

**Fig. 3**

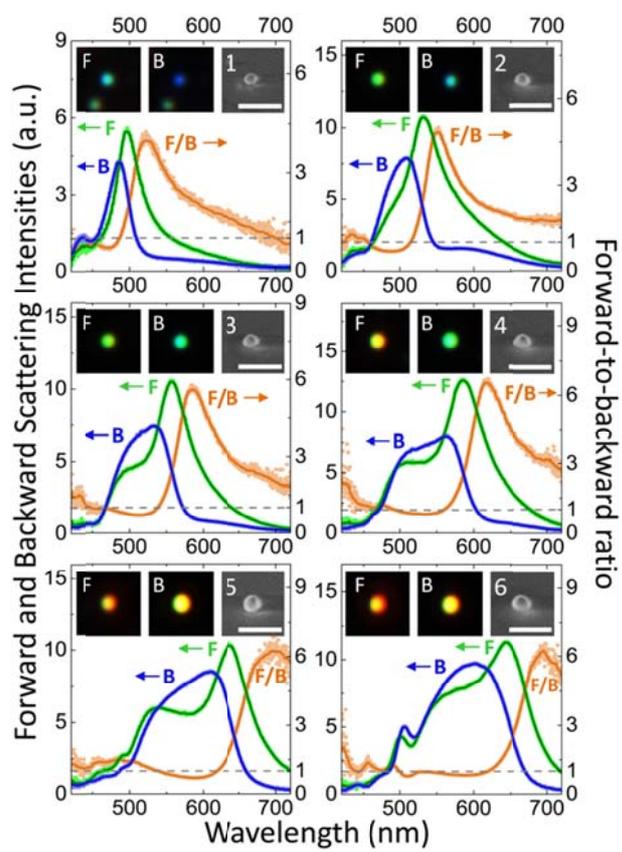

**Fig. 4**

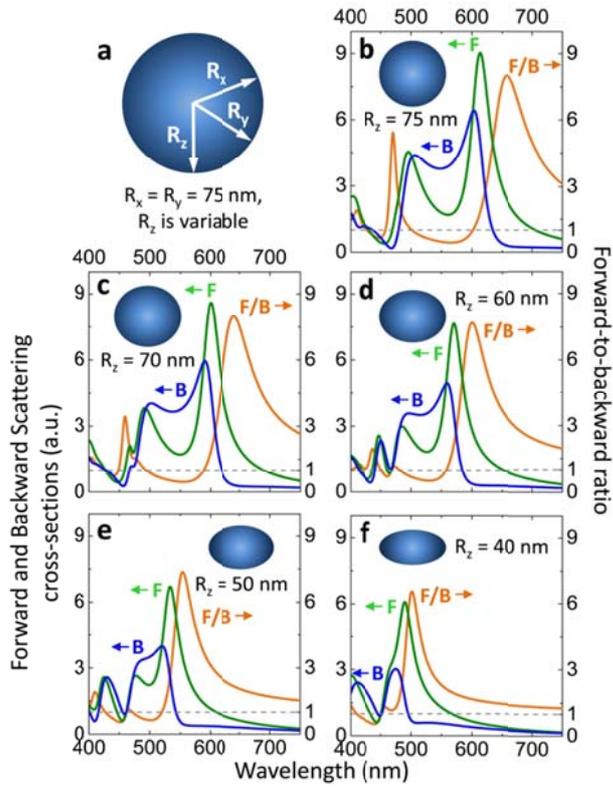

# Supplementary Information

# Directional visible light scattering by silicon nanoparticles

*Yuan Hsing Fu,[1] Arseniy I. Kuznetsov,[1,] * Andrey E. Miroshnichenko,[2] Ye Feng Yu,[1] and Boris Luk'yanchuk,[1]*

1. Data Storage Institute, 5 Engineering Drive 1, 117608, Singapore
2. Nonlinear Physics Centre, Research School of Physics and Engineering, the Australian National University, Canberra, 0200, Australia

*Corresponding author:  arseniy_k@dsi.a-star.edu.sg

Figure S1 shows detailed illustration of the forward and backward scattering measurements. Numerical aperture (NA) of the upper dark-field condenser is of 0.8-0.9 as shown in Figure S1a. Hence, the angle of light incidence onto the nanoparticle sample varies from 53º to 64º with average value of 58.5º with respect to the normal to the surface. The NA of the bottom objective lens for light collection is of 0.55, which means the total angle for light collection is of 67º. If we now consider the light propagation direction along the red line, which is marked by signs 0º and 180º in Figure S1a, the collection angle for light scattering will be from 25º to 92º relative to this direction. Taking into account rotational symmetry this will correspond to about a half of a total forward scattering of the particle. Backward scattering is not collected in these conditions. Similar situation is realized in the backward scattering measurements as shown in Figure S1b. The objective lens collects the scattered light into the angle from 88º to 155º relative to the light propagation direction. This corresponds to about a half of the total backward scattering without the forward scattering contribution. As a result, the forward and backward scatterings are measured separately.

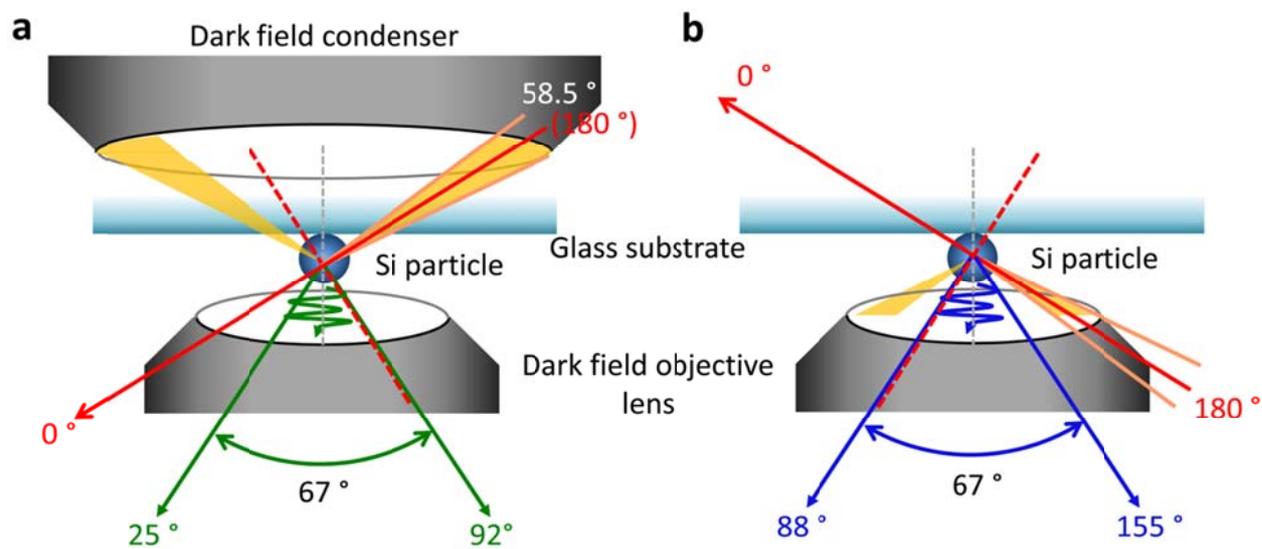

**Figure S1.** Detailed schematic illustration of the forward **(a)** and backward **(b)** scattering measurements.